\documentclass[pra,twocolumn,aps,floatfix,showpacs,tightenlines,showpacs,superscriptaddress]{revtex4}
\usepackage{amssymb,amsmath,amsthm,amsbsy,epsfig,color,graphicx,times}

\begin{document}

\title{Tunable non-Gaussian resources for continuous-variable quantum technologies}
\author{F. Dell'Anno}
\affiliation{Liceo Statale P. E. Imbriani, via Pescatori 155, 83100 Avellino, Italy}
\author{D. Buono}
\affiliation{Dipartimento di Ingegneria Industriale, Universit\`{a} degli Studi di
Salerno, via Giovanni Paolo II, I-84084 Fisciano (SA), Italy}
\affiliation{CNISM - Consorzio Nazionale Interuniversitario per le Scienze Fisiche della Materia, Unit\`a di Salerno, I-84084 Fisciano (SA), Italy}
\author{G. Nocerino}
\affiliation{Trenitalia spa, DPR Campania, Ufficio di Ingegneria della Manutenzione, IMC
Campi Flegrei, Via Diocleziano 255, 80124 Napoli, Italy}
\author{A. Porzio}
\email{alberto.porzio@spin.cnr.it}
\affiliation{CNR -- SPIN, Unit\`{a} di Napoli, Complesso Universitario Monte Sant'Angelo,
I-80126 Napoli, Italy}
\author{S. Solimeno}
\affiliation{Dipartimento di Scienze Fisiche, Universit\`{a} \textquotedblleft Federico II",
Complesso Universitario Monte Sant'Angelo, I-80126 Napoli, Italy}
\author{S. De Siena}
\email{desiena@sa.infn.it }
\affiliation{Dipartimento di Ingegneria Industriale, Universit\`{a} degli Studi di
Salerno, via Giovanni Paolo II, I-84084 Fisciano (SA), Italy}
\affiliation{CNISM - Consorzio Nazionale Interuniversitario per le Scienze Fisiche della Materia, Unit\`a di Salerno, I-84084 Fisciano (SA), Italy}
\author{F. Illuminati}
\email{illuminati@sa.infn.it }
\affiliation{Dipartimento di Ingegneria Industriale, Universit\`{a} degli Studi di
Salerno, via Giovanni Paolo II, I-84084 Fisciano (SA), Italy}
\affiliation{CNISM - Consorzio Nazionale Interuniversitario per le Scienze Fisiche della Materia, Unit\`a di Salerno, I-84084 Fisciano (SA), Italy}
\date{August 9, 2013}

\begin{abstract}
We introduce and discuss a set of tunable two-mode states of
continuous-variable systems, as well as an efficient scheme for their
experimental generation. This novel class of tunable entangled resources is
defined by a general ansatz depending on two experimentally adjustable
parameters. It is very ample and flexible as it encompasses Gaussian as well
as non-Gaussian states. The latter include, among others, known states such
as squeezed number states and de-Gaussified photon-added and photon-subtracted
squeezed states, the latter being the most efficient non-Gaussian resources
currently available in the laboratory. Moreover, it contains the classes of
squeezed Bell states and even more general non-Gaussian resources
that can be optimized according to the specific quantum technological
task that needs to be realized. The proposed experimental scheme exploits
linear optical operations and photon detections performed on a pair of
uncorrelated two--mode Gaussian squeezed states. The desired non-Gaussian
state is then realized via ancillary squeezing and conditioning. Two
independent, freely tunable experimental parameters can be exploited to
generate different states and to optimize the performance in implementing a
given quantum protocol. As a concrete instance, we analyze in detail the
performance of different states considered as resources for the
realization of quantum teleportation in realistic conditions.
For the fidelity of teleportation of an unknown coherent
state, we show that the resources associated to the optimized parameters
outperform, in a significant range of experimental values, both Gaussian
twin beams and photon-subtracted squeezed states.
\end{abstract}


\maketitle







\section{Introduction}

Quantum information with Gaussian states of continuous variable systems has
been investigated thoroughly and with startling success both theoretically
and experimentally (for comprehensive reviews of different aspects see,
e.g., Refs.~\cite{Adesso2007,Braunstein2005,Kok2007,Eisert2003,Dellanno2006}).
At the same time, there has been growing awareness of some important
limitations intrinsic to working entirely within the framework of Gaussian
states and/or Gaussian operations, so that at some point it becomes both
desirable and necessary to start exploring the vast ocean of
continuous-variable non-Gaussian states and non-Gaussian operations. The
first pioneering and paradigmatic example of this need is given by the by
now classic proof that distillation of Gaussian entangled states resorting
only to Gaussian operations is impossible~\cite{Scheel2002}.

In fact, various compelling reasons suggest a thorough investigation of the
properties of continuous-variable non-Gaussian states. Gaussian states are
indeed extremal in the sense that at fixed covariance matrix several
nonclassical properties such as entanglement, when measured by the entanglement
of formation, and distillable secret key rates are minimized by Gaussian
states~\cite{Wolf2006}. Specific families of non-Gaussian entangled resources
lead to significant improvements in the performance of existing continuous-variable
quantum information protocols such as teleportation~\cite{Dellanno2007,Dellanno2010}.
Measurement-based quantum computation with continuous variables eventually needs both
non-Gaussian operations and non-Gaussian resource states in order to be
universal~\cite{Ohliger2010,Ohliger2012}. Preliminary investigations of the
interplay of non-Gaussian inputs and different types of Gaussian and
non-Gaussian channels unveils the clear advantages, in general, of
non-Gaussianity for state estimation and metrology~\cite{Adesso2009,Monras2010,Monras2011,Datta2012,Genoni2013}. Stronger violations
of Bell inequalities and better performing entanglement swapping protocols
are also expected using non-Gaussian resources and strategies beyond the
currently available ones~\cite{Obermaier2011,Banaszek1999,Son2009,Sciarrino2012,Sciarrino2013}.
While so-called Gaussifier protocols of entanglement distillation have been
proposed that convert noisy non-Gaussian states into Gaussian ones through
converging iterative procedures~\cite{Distillery2012,Campbell2013}, general
optimal strategies for the distillation of highly entangled non-Gaussian
states are still lacking. Finally, the interplay of non-Gaussianity,
non-classicality, and non-Markovianity is expected to lead to further new
phenomena when addressing the dynamics of open continuous-variable systems.

Among the many families of continuous-variable non-Gaussian states that can
be of interest for fundamental quantum physics as well as for quantum
information and related quantum technologies one should mention squeezed cat
states~\cite{Dellanno2006}; multiphoton squeezed states, that is states that
generalize the usual two-photon Gaussian squeezed states by considering
nonlinear extensions of the linear Bogoliubov squeezing transformations,
either single-mode~\cite{Dellanno2004} or two-mode~\cite{Dellanno2004bis};
and especially squeezed Bell states~\cite{Dellanno2007}. Indeed, the latter include
squeezed Fock and de-Gaussified squeezed coherent states as particular
cases, as we shall review below. Non-Gaussian and De-Gaussified states can
be generated either by introducing higher-order nonlinearities in the
source, and/or by performing conditional measurements.

Many theoretical and experimental efforts have been concentrated on the
engineering of nonclassical, non-Gaussian states of the radiation
field~\cite{Za,We,Ta,Our,Lv,Da,dauria05}. In particular, concerning quantum
teleportation with continuous variables~\cite{Vaidman,BrauKimb,Furusawa98,Lee11},
it has been demonstrated that the fidelity of teleportation can be improved using
various families of non-Gaussian resources~\cite{Opa,Co,Oli}. At present, the best
experimentally realized non-Gaussian resource for continuous-variable
teleportation is the photon-subtracted squeezed state~\cite{We,Ta,Our}. On
the other hand, as already mentioned, a new class of continuous-variable
non-Gaussian states, the Squeezed Bell states, has been introduced
recently~\cite{Dellanno2007}. These states interpolate between different
de-Gaussified states, and can be fine tuned by acting on an independent free
parameter in addition to the squeezing.

From a theoretical point of view, entangled Gaussian and de-Gaussified
states are defined by applying squeezing and ladder operators on the
two-mode vacuum. Within this theoretical context the teleportation fidelity
for the Braunstein-Kimble-Vaidman protocol is improved, in a significant
range of the parameters, by replacing Gaussian and de-Gaussified resources
by optimized squeezed Bell states~\cite{Dellanno2007}. This has been
verified for different inputs (including coherent states, squeezed states,
and squeezed number states), also in the presence of losses and other
realistic sources of imperfections~\cite{Dellanno2010}. This effect can be
understood by remarking a crucial difference existing between the case of
teleportation protocols relying on entangled Gaussian resources and the case
allowing for more general entangled states. In the former it is known that
the fidelity of teleportation and the entanglement of the shared entangled
Gaussian resource are in one-to-one correspondence~\cite{Adesso2005}. In the
latter this is no longer true and the fidelity of teleportation becomes a
highly complicated function of three (in general conflicting) variables:
degree of entanglement, degree of non-Gaussianity, and degree of Gaussian
affinity~\cite{Dellanno2007}. In particular, the third variable (Gaussian
affinity) is crucial. It quantifies the overlap with the two-mode squeezed
vacuum. Loosely speaking, it assures that an efficient entangled resource
must contain a contribution, with a relevant weight, given by the two-mode
squeezed vacuum plus symmetric non Gaussian corrections. The optimized
squeezed Bell states realize indeed the best possible compromise for the
simultaneous maximization over all these three properties of the fidelity of
teleportation~\cite{Dellanno2007}.

Going beyond these preliminary theoretical results, the desired goal would
be to construct experimental platforms capable of generating classes of
highly tunable non-Gaussian resources with enhanced performances for
protocols of quantum technology based on continuous variables. In order to
proceed towards concrete experimental realizations, one needs to introduce a
basic scheme of generation that takes into account all the relevant sources
of noise and imperfections in realistic instances. Thereafter, one must
verify that in these realistic scenarios the performance of the new resource
in the framework of a given quantum protocol provides an appreciable
advancement that justifies the experimental effort. After this preliminary
analysis, one needs to work out carefully the details of the experimental
setup and, finally, one needs to provide reliable methods for the
reconstruction of the experimentally generated states.

In the present work we introduce the basic scheme of generation for a large
class of tunable two-mode entangled non-Gaussian states and we present a
preliminary analysis on their performance as resources for
continuous-variable quantum technologies. The experimental scheme that we are
going to introduce has the advantage of being flexible and versatile, in the sense
that a variation of the freely adjustable experimental parameters allows for the
generation of different non-Gaussian states including, besides the squeezed
Bell states, photon-added squeezed states~\cite{Li02}, photon-subtracted squeezed
states~\cite{Ta}, squeezed number states and, obviously, also Gaussian twin
beams~\cite{dauria09,Buono10}. The relative performance of different states
will will be investigated in detail for the protocol of quantum
teleportation of unknown coherent states~\cite{Vaidman,BrauKimb}. We will
show that the optimized states in realistic conditions provide, in a
significant range of physical parameters, a superior performance compared to all
existing de-Gaussified states.

\section{Experimental scheme with adjustable free parameters for the
generation of non-Gaussian states}

The pure (normalized) squeezed Bell states originally introduced in
Refs.~\cite{Dellanno2007,Dellanno2010} are of the form
\begin{equation}
|\Psi >_{\text{\textbf{SB}}}\equiv S_{12}\left( -r\right) \left\{ \cos
\delta |0,0>_{12}+\sin \delta |1,1>_{12}\right\} \; ,  \label{SB}
\end{equation}
where $|0,0>_{12}$ and $|1,1>_{12}$ denote, respectively, the tensor product
of two single-mode vacua and two one-photon states, while $S_{ij}\left(
-r\right) = \exp \left\{ ra_{i}^{\dag }a_{j}^{\dag
}-ra_{i}a_{j}\right\}$ is the
two-mode squeezing operator, and $\delta $ is a free parameter allowing for
optimization. A more general form of the squeezed Bell states could include
a relative phase, but this inclusion would not improve the performance of
the squeeze Bell states when considered as entangled resources. At some
suitably chosen values of the $\delta $ parameter, the squeezed Bell
superposition coincides with de-Gaussified photon-added states,
photon-subtracted states, with squeezed number states, and with Gaussian
twin beams~\cite{Dellanno2007}, where addition/subtraction operations, as
well as the number state, are referred to the case of a single photon. For a
reminder, we list in Table~\ref{TheorDef} the theoretical definitions of all
the states considered.

\begin{table}[tp]
\begin{tabular}{|c|c|}
\hline
\textbf{State} & \multicolumn{1}{||c|}{\textbf{Definition}} \\ \hline\hline
Photon-subtracted squeezed state & $N_{PS}\;a_{1}a_{2}S_{12}(\zeta
)|0,0>_{12}$ \\ \hline
Photon-added squeezed state & $N_{PA}\;a_{1}^{\dagger }a_{2}^{\dagger
}S_{12}(\zeta )|0,0>_{12}$ \\ \hline
Squeezed number state & $S_{12}(\zeta )|1,1>_{12}$ \\ \hline
Gaussian twin beam & $S_{12}(\zeta )|0,0>_{12}$ \\ \hline
\end{tabular}
\caption{Theoretical definition of some states of particular interest that
are included in the class of squeezed Bell states. $N_{PS}$ and $N_{PA}$
denote, respectively, the normalization of de-Gaussified photon-subtracted
and photon-added squeezed states.}
\label{TheorDef}
\end{table}

\begin{figure}[ht]
\includegraphics*[width=8cm]{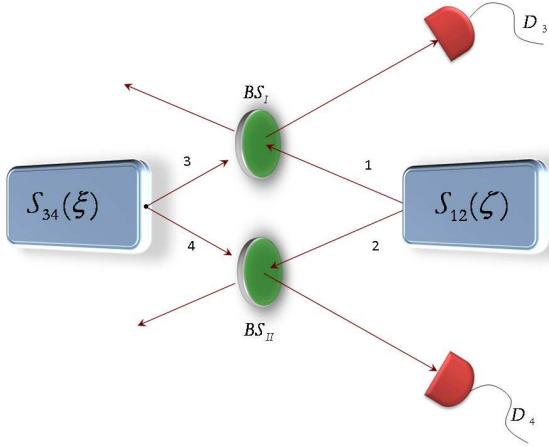}
\caption{Sketch of the ideal scheme for the experimental generation of the
states defined theoretically in Eq.~(\protect\ref{psi_12}). Two
Gaussian twin beams, generated independently, respectively
$\left\vert \protect\zeta \right\rangle _{12}$ and $\left\vert \protect\xi \right\rangle _{34}$,
impinge on two beam splitters $BS_{I}$ and $BS_{II}$ of
transmissivity $T_{1}$ and $T_{2}$. The generation of the output two-mode
state is triggered by two simultaneous detections realized by the
single-photon projective detectors $D_{3}$ and $D_{4}$.}
\label{FigSchemeIdeal}
\end{figure}

In this section we introduce a scheme capable to generate two-mode
non-Gaussian states of the electromagnetic field that provide the best
experimental approximation to the form and/or to the performance of the
theoretically defined squeezed Bell states. The idea is to manipulate an
overall four-mode system described by two independent Gaussian twin beams,
one of which will play the role of an ancillary two-mode state, and then
exploit linear optical components and conditional measurements. In this
respect, we recall that twin beams are routinely generated in type II
Optical Parametric Oscillators (OPOs)~\cite{Fornaro08}. Therefore, in
our scheme, no higher-order nonlinearities are either needed or desirable.
The basic generation scheme is illustrated in
Fig.~(\ref{FigSchemeIdeal}).

In this scheme we exploit two independent Gaussian twin beams,
$|\zeta >_{12}=S_{12}(\zeta )|0,0>_{12}$ and
$|\xi >_{34}=S_{34}(\xi)|0,0>_{34}$, so that we start with an initial
four-mode "proto-state"
\begin{equation}
|\zeta >_{12}|\xi >_{34}=S_{12}(\zeta )S_{34}(\xi )|\mathbf{0}>_{1234} \; ,
\label{protos}
\end{equation}
where $|\mathbf{0}>_{1...n}=\bigotimes\nolimits_{k=1}^{n}|0>_{k}$ denotes
the tensor product of $n$ single-mode vacuum states. The twin beams feed the
input ports of two beam splitters, respectively of transmissivity $T_{1}$
and $T_{2}$. Modes $1,3$ mix at the beam splitter $BS_{I}$, while modes $2,4$ mix at the beam
splitter $BS_{II}$. The resulting state is a four-mode entangled state
$|\Phi >_{1234}$ of the form
\begin{eqnarray}
&&|\Phi >_{1234}=U_{13}(\kappa _{1})U_{24}(\kappa _{2})|\zeta >_{12}|\xi>_{34}  \notag \\
&=&U_{13}(\kappa _{1})U_{24}(\kappa _{2})S_{12}(\zeta )S_{34}(\xi )
|\mathbf{0}>_{1234} \, ,
\label{phi_BS}
\end{eqnarray}
where $S_{12}(\zeta )$
and $S_{34}(\xi )$ are the squeezing operators with complex squeezing
parameters $\zeta =r\exp \left\{ i~\phi _{\zeta }\right\} $ and $\xi =s\exp
\left\{ i~\phi _{\xi }\right\} $ respectively. The beam splitter operators
read $U_{lk}(\kappa _{l})=\exp \left\{ \kappa _{l}\left( a_{l}^{\dag
}a_{k}-a_{l}a_{k}^{\dag }\right) \right\} $, where $l=1,k=2$ for the first
beam splitter, and $l=3,k=4$ for the second one. Finally, $\tan \kappa _{l}=
\sqrt{\left( 1-T_{l}\right) /T_{l}}$.

Starting with the four-mode state $|\Phi >_{1234}$, the conditional
measurements provided by the simultaneous clicks of the detectors
$D_{3},D_{4}$, and the restriction to suitable ranges of the beam splitters
parameters and of the squeezing parameters, will lead to the generation of
two-mode states which, as we will discuss, provide an approximate
realization of the theoretical squeezed Bell states Eq. (\ref{SB}).
Obviously, the experimental generation
implies non ideal conditions, including losses and detection inefficiency.
We will proceed in steps. We will consider the ideal situation first, with
perfect \emph{single-photon conditional measurements}. This first step is
useful in order to describe the basic elements of the scheme and the
connection to the theoretical squeezed Bell states. In a second step, we
will discuss the full realistic instance: we will include losses and detection
inefficiency, with numerical figures well within the range of those
accessible in current experiments.

\subsection{Single-photon conditional measurements}

Here we will consider detectors that are perfectly photon-resolving with
perfect coincidence in the simultaneous detections of single photons in
modes $3$ and $4$. Within this idealization, simultaneous detections project
the state of Eq.~(\ref{phi_BS}) onto the \textit{tunable state}
$|\Psi_{\text{\textbf{T}}} \rangle$:
\begin{equation}
|\Psi_{\text{\textbf{T}}} \rangle = {\mathcal{N}}_{34} \langle
1,1|U_{13}(\kappa _{1})U_{24}(\kappa_{2}) S_{12}(\zeta) S_{34}(\xi)
|\mathbf{0} {\rangle}_{1234} \; ,
\label{psi_12}
\end{equation}
where $\mathcal{N}$ denotes the normalization constant.

Varying the six free parameters, $\kappa _{1}$, $\kappa _{2}$, $r$, $s$,
$\phi _{\zeta }$, $\phi _{\xi }$ the setup can produce different two-mode
states: fully non-Gaussian, de-Gaussified, and Gaussian. Let us start by
fixing $\phi _{\zeta }=\pi $; then $S_{12}(\zeta )\equiv S_{12}(-r)$.
On the other hand, if we
fix also $\phi _{\xi }=\pi $ (as it indeed will be later forced by
optimization, see next section) it is straightforward to realize that the
two beam splitters become indistinguishable:
\begin{equation*}
T_{1}=T_{2}\equiv T\; ,
\end{equation*}
and thus
\begin{equation*}
\kappa _{1},\kappa _{2}\equiv \kappa \; .
\end{equation*}
Moreover, it is immediate to see that $\xi =-s$ and $S_{34}(\xi )\equiv
S_{34}(-s)$. This simplified instance is sufficient for the purpose of
generating the general class of squeezed Bell states. Further, we consider
the situation in which $\kappa ^{2}<<1$, while the amplitude $|\xi |(\equiv
s)$ of the ancillary squeezing $S_{34}$ is chosen to be at most of the same
order of $\kappa ^{2}$ (the significance of this choice will be clarified
below). Therefore, we are considering beam splitters with a high
transmissivity $T=\cos ^{2}|k|$, and an ancillary squeezing $S_{34}$ with a
weak relative squeezing amplitude. As a consequence of these choices, the
unitary operators $U_{13}(\kappa ),\;U_{24}(\kappa )$ can be expanded in a
power series truncated at the order $\kappa ^{2}$, while $S_{34}(\xi )$ can
be truncated at the order $|\xi |\equiv s$. Wrapping up, under these
conditions one finds that
\begin{eqnarray}
&&|\Phi \rangle _{1234}\approx  \notag  \label{psifourmodeBSex} \\
&&\left[ 1+\kappa (a_{1}^{\dag }a_{3}-a_{1}a_{3}^{\dag })+\frac{\kappa
^{2}(a_{1}^{\dag }a_{3}-a_{1}a_{3}^{\dag })^{2}}{2}+\mathcal{O}(\kappa ^{3})
\right] \times  \notag \\
&&\left[ 1+\kappa (a_{2}^{\dag }a_{4}-a_{2}a_{4}^{\dag })+\frac{\kappa
^{2}(a_{2}^{\dag }a_{4}-a_{2}a_{4}^{\dag })^{2}}{2}+\mathcal{O}(\kappa ^{3})
\right] \times  \notag \\
&&\left[ 1+(s(a_{3}^{\dag }a_{4}^{\dag }-a_{3}a_{4})+\mathcal{O}(s^{2}))
\right] \times  \notag \\
&&S_{12}(-r)|0,0,0,0\rangle _{1234} \; .
\end{eqnarray}
Next, we apply a postelection strategy (see Appendix A for
more details). By using photo-detection in
coincidence, the conditional measurements of simultaneous detections of
single photons in modes $3$ and $4$ project the non-normalized state
Eq.~(\ref{psifourmodeBSex}) onto the reduced overlap two-mode state
$\,_{34}<1,1|\Phi >_{1234}$:
\begin{equation}
\,_{34}<1,1|\Phi >_{1234} \approx
(s + \kappa^{2}a_{1}a_{2})S_{12}(-r)|0,0>_{12} \; .
\label{postselectSB}
\end{equation}
Due to our assumptions on the relative amplitude of the parameters
$\kappa^{2}$ and $|\xi |$, in the above equation we have neglected terms
proportional to $|\xi |\kappa ^{2}$, that is all contributions of the form
$-s\kappa ^{2}(a_{1}^{\dag }a_{1}+a_{2}^{\dag }a_{2})S_{12}(-r)|0,0>_{12}$,
as well as all those of higher order. Exploiting the two-mode Bogoliubov
transformations
\begin{eqnarray}
&&S_{12}^{\dag }(-r)\,a_{i}\,S_{12}(-r)=\cosh r\,a_{i}+\sinh r\,a_{j}^{\dag} \, ,  \notag \\
&&(i\neq j=1,2) \; ,
\label{BogoliubovT}
\end{eqnarray}
we finally obtain the non-normalized two-mode state
\begin{eqnarray}
&&S_{12}(-r)\,\left\{ (s+\kappa ^{2}\sinh r\cosh r)|0,0\rangle _{12}\right.
\notag \\
&&\left. +\kappa ^{2}\sinh ^{2}r|1,1>_{12}\right\} \; ,
\label{expSqueezedBell}
\end{eqnarray}
whose form, apart from normalization, coincides with that of the theoretical
squeezed Bell state Eq.~(\ref{SB}). Normalizing, we obtain finally
\begin{eqnarray}
&&|\psi _{\text{\textbf{T}}}>_{12}=S_{12}(-r) \,
\left\{c_{00}|0,0>_{12}+c_{11}|1,1>_{12}\right\} \, ,
\label{expSqueezedBell2} \\
&&  \notag \\
&&c_{00}=\frac{-\lambda +\sinh r\cosh r}{[(-\lambda +\sinh r\cosh
r)^{2}+(\sinh ^{2}r)^{2}]^{1/2}} \, ,  \label{c00} \\
&&c_{11}=(1-c_{00}^{2})^{1/2} \; ,  \label{c11}
\end{eqnarray}
where $\lambda =-s/\kappa ^{2}$. The form Eq.~(\ref{SB}) is recovered
observing that
\begin{equation}
\delta =\arctan \left( \frac{\kappa ^{2}\sinh ^{2}r}{s+\kappa ^{2}\sinh
r\cosh r}\right) \; .
\label{deltasper}
\end{equation}
Photon-added and photon-subtracted squeezed states, squeezed number states, twin
beams, and other particular states in the class Eq.~(\ref{expSqueezedBell2})
can then be obtained by choosing the experimental parameters in such a way
that the free parameter $\delta $, see Eq.~(\ref{deltasper}), takes the
corresponding special values~\cite{Dellanno2007}. Including terms of higher
order in the expansion, the ensuing family of states still realizes close approximations
to the theoretical squeezed Bell states. For instance, suppose that one
truncates at order $\kappa^{4}$ in the beam splitter operators. By
consistency, one needs then to truncate at order $|\xi |^{3}$ in the
squeezing operator. By imposing the constraint that $|\xi |^{2}$ be at most
of order $\kappa^{3}$, one recovers again the squeezed Bell states with the
same rate of approximation. Therefore, one is always justified in
considering only truncations at lowest order in $\kappa ^{2}$.

The discussion of the ideal experimental setup allows a clear understanding
of the general idea, by showing that the basic scheme can generate, in a
controlled manner, states arbitrarily close to the theoretical squeezed Bell
states. On the other hand, we can relax to some extent the constraint that
the shape of the generated states be exactly that of the squeezed Bell
states, given that the main aim is to generate states with enhanced
performances with respect to Gaussian twin beams and de-Gaussified squeezed
states. Therefore, in the subsequent analysis of realistic schemes, while
retaining the condition $\kappa^2 << 1$, we will allow $s$ to vary
arbitrarily.

\subsection{Generation under realistic conditions}

In realistic experimental conditions the state
$|\Psi _{\text{\textbf{T}}}\rangle $ will be affected by unavoidable sources of decoherence such as
cavity output couplings and losses during propagation~\cite{DAuria06,Buono12}.

In this context, the four-mode proto-state $|\zeta>_{12} |\xi {\rangle}_{34}$,
Eq.~(\ref{protos}), turns into a four-mode squeezed thermal state
described by the following input density matrix (see appendix B for
details):
\begin{equation}
\rho_{1234}=S_{12}\left( \zeta \right) S_{34}\left( \xi \right)
\rho_{1234}^{th} S_{12}^{\dag }\left( \zeta \right) S_{34}^{\dag }
\left( \xi \right) \; ,
\end{equation}
where $\rho_{1234}^{th}=\bigotimes_{k=1}^{4} \rho _{k}^{th}$ and $\rho_{k}^{th}$
is the density matrix of the thermal state associated to mode
$k$. On the other hand, at typical room temperatures, the thermal density
matrix $\rho_{1234}^{th}$ tends to the vacuum state, so that $\rho_{1234}$
coincides for all practical purposes with the projection operator associated
to the pure state $|\Phi>_{1234}$ (see Appendix B for details).

\begin{figure}[ht]
\includegraphics[width=9cm]{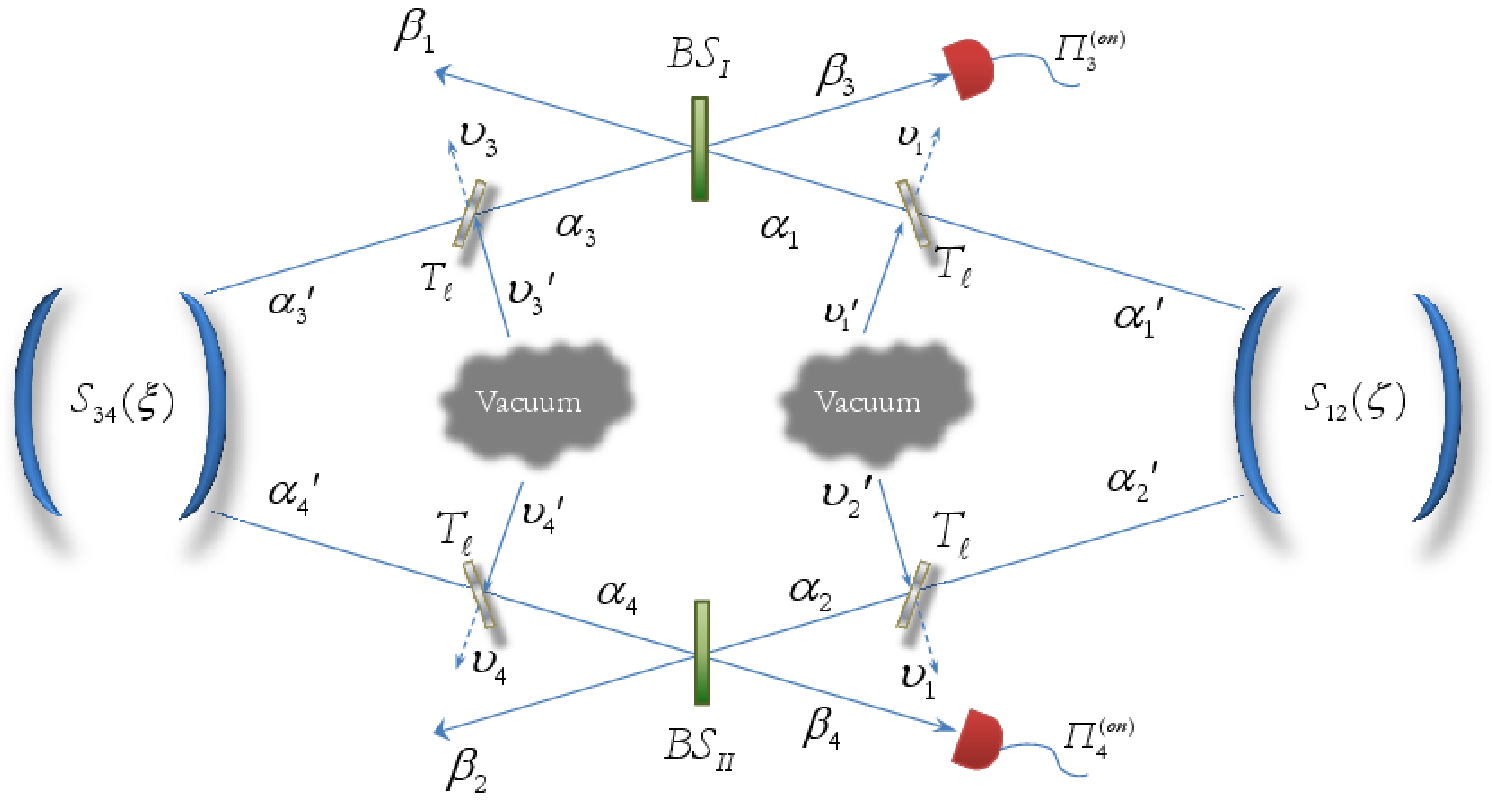}
\caption{Scheme of generation of tunable two-mode states under realistic conditions:
two independently produced Gaussian twin beams, $\left\vert \protect\zeta \right\rangle _{12}$
and $\left\vert \protect\xi \right\rangle _{34}$, are mixed at the two beam
splitters $BS_{I}$ and $BS_{II}$ of transmissivity $T_{1}$ and $T_{2}$. Four
fictitious beam splitters with transmissivity $T_{\ell }$ mimic the
various decoherence mechanisms. The single-photon projective measurements
are replaced by the positive-operator valued measures (POVMs) $\Pi _{3}^{(on)}$ and $\Pi _{4}^{(on)}$ with
quantum efficiencies $\protect\eta <1$.}
\label{SchemeReal}
\end{figure}

In Fig.~(\ref{SchemeReal}) we illustrate the scheme of generation under
realistic conditions. The decoherence mechanisms are modeled by introducing
four fictitious beam splitters (one for every input mode) with equal
transmissivity $T_{\ell }\left( =1-R_{\ell }\right) $. Each beam splitter is
illuminated at the empty port by a single-mode vacuum $\upsilon _{k}$. As
already mentioned, at room temperature the thermal contribution is
negligible, and thus one simply needs to replace the state $|\Phi >_{1234}$
of Eq.~(\ref{phi_BS}) with the state
\begin{equation}
|\Phi ^{\prime }>_{1234}=\bigotimes\limits_{k=1}^{4}U_{k}\left( T_{\ell
}\right) |\Phi >_{1234} \; ,
\label{phi_loss}
\end{equation}
where the beam splitter operator that mixes mode $a_{k}$ with the vacuum $v_{k}$ is given by
$U_{k}\left( T_{\ell }\right) =\exp \left\{ \kappa _{\ell
}a_{k}^{\dag }v_{k}-\kappa _{\ell }^{\ast }a_{k}v_{k}^{\dag }\right\} $, and
$\kappa _{\ell }$ is such that $\tan \kappa _{\ell }=\sqrt{\left( 1-T_{\ell
}\right) /T_{\ell }}$. The postselection procedure is implemented as follows
(see appendix A for further details). The detection associated to modes
$k=3,4$ is now modeled by the positive-operator valued measure (POVM) $\Pi _{k}^{(on)}(\eta _{k})$ that takes into account the threshold detection of $n\geq 1$ photons:
\begin{equation}
\Pi _{k}^{(on)}(\eta _{k})=\mathbb{I}_{k}-\Pi _{k}^{(off)}(\eta _{k}) \; ,
\label{PiOn}
\end{equation}
where
\begin{equation}
\Pi _{k}^{(off)}(\eta _{k})=\sum_{m=0}^{\infty }\left( 1-\eta _{k}\right)
^{m}|m>_{k\text{ \ }k}<m| \; ,
\label{PiOff}
\end{equation}
and $\eta _{k}$ is the non-unit detection efficiency for mode $k$. The
corresponding density matrix reads
\begin{equation}
\rho _{\text{\textbf{T}}}^{(on)}\left( T_{\ell },\eta _{3},\eta _{4}\right) =
\frac{\text{Tr}_{34}\left[ \rho _{1234}^{\prime } \otimes \Pi_{3}^{(on)}(\eta _{k})
\otimes \Pi _{4}^{(on)}(\eta _{k})\right]}{\mathcal{N}_{\text{\textbf{T}}}^{(on)}
\left( \eta _{3},\eta _{4}\right)} \; ,
\label{POVMid}
\end{equation}
where $\rho _{1234}^{\prime }$ is the density matrix relative to the state
$|\Phi ^{\prime }>_{1234\text{ }}$. The normalization constant reads
\begin{equation}
\mathcal{N}_{\text{\textbf{T}}}^{(on)}\left( \eta _{3},\eta _{4}\right) =
\text{Tr}_{1234}\left[ \rho _{1234}\otimes \Pi _{3}^{(on)}(\eta _{k})\otimes
\Pi _{4}^{(on)}(\eta _{k})\right] \; .
\label{N_T_on}
\end{equation}
It depends on $\eta _{3},\eta _{4}$ and represents the success rate for
entanglement distillation in a realistic scenario~\cite{vanLock2011}. In the
presence of losses $\left( T_{\ell }<1\right) $ and of imperfect quantum
efficiencies ($\eta _{3},\eta _{4}<1$), the corresponding approximations to
squeezed Bell states, photon-subtracted squeezed states, and Gaussian twin
beams are obtained by inserting the values of the ancillary parameters that
yield these states in the theoretical instance~\cite{Dellanno2007}:
\begin{eqnarray*}
\rho _{\text{\textbf{PS}}}^{(on)}\left( T_{\ell },\eta _{3},\eta _{4}\right)
&=&\left. \rho _{\text{\textbf{T}}}^{(on)}\left( T_{\ell },\eta _{3},\eta
_{4}\right) \right\vert _{s=0,\kappa \simeq 0}, \\
\rho _{\text{\textbf{SB}}}^{(on)}\left( T_{\ell },\eta _{3},\eta _{4}\right)
&=&\left. \rho _{\text{\textbf{T}}}^{(on)}\left( T_{\ell },\eta _{3},\eta
_{4}\right) \right\vert _{s\simeq \kappa ^{2}\ll 1,\phi =\pi }, \\
\rho _{\text{\textbf{TB}}}^{(on)}\left( T_{\ell },\eta _{3},\eta _{4}\right)
&=&\left. \rho _{\text{\textbf{T}}}^{(on)}\left( T_{\ell },\eta _{3},\eta
_{4}\right) \right\vert _{\xi =\zeta \equiv \varepsilon } \; .
\end{eqnarray*}

A further practical restriction, due to decoherence, comes from the fact
that the effective value of the squeezing parameters is reduced. In Appendix
A we show in detail that the actual squeezing parameter $r^{\prime }$ is
related to the loss-free ideal parameter $r$ according to the following
correspondence:
\begin{equation}
r^{\prime }=-\frac{1}{2}\ln \left[ 1-T_{\ell }\left( 1-e^{-2r}
\right) \right] \; .
\end{equation}
For instance, if in the
block scheme of Fig.~(\ref{FigSchemeIdeal}) the squeezing is fixed at $r=2$
($\simeq 17.4$ dB), in realistic conditions, with a $15\%$ level of losses
($T_{\ell }=0.85$), it corresponds to a beam with $r^{\prime }$ of about $0.90$
($\simeq 7.81$ dB).

In the following, we will discuss the performance of non-Gaussian entangled
resources in implementing quantum teleportation protocols, as measured by
the teleportation fidelity. Given that the photon-added squeezed states and
the squeezed number states, due to their very low degree of Gaussian affinity,
can never outperform Gaussian twin beams with the same covariance matrix,
as already discussed, e.g., in Ref.~\cite{Dellanno2007},
in the following we will compare optimized squeezed Bell
states, photon-subtracted squeezed states, and Gaussian twin beams.

\section{Tunable non-Gaussian resources and quantum teleportation}

\textit{Preliminaries --} In this Section we seek to optimize the fidelity
of the Braunstein-Kimble-Vaidman teleportation protocol of unknown coherent
states~\cite{Vaidman,BrauKimb} using, as two-mode entangled resources, the
states generated using the realistic scheme introduced in the previous
section. To this end, it is convenient to exploit the formalism of the
characteristic function~\cite{Marian}, which is particularly suited for the
analysis of non-Gaussian states, because it greatly simplifies the
computational strategies~\cite{Dellanno2007}.

For an $n$-mode state described by a density matrix $\rho $ the
characteristic function is defined as
\begin{equation}
\chi (\beta _{1},...,\beta _{n})=\text{Tr}[\rho D_{1}(\beta _{1})\otimes
...\otimes D_{n}(\beta _{n})] \; ,
\label{chidef}
\end{equation}
where $D_{i}(\beta _{i})$ denotes the Glauber displacement operator for the
mode $i$ ($i=1,...,n$).

In Appendix A we show in detail
that, given a four--mode state represented by the characteristic function
$\chi _{1234}\left( \beta _{1};\beta _{2};\beta _{3};\beta _{4}\right)$,
the state achieved after conditional measurements on the two ancillary modes $3$
and $4$, see Fig.~(\ref{FigSchemeIdeal}), is given by the characteristic
function
\begin{eqnarray}
&&\chi _{\text{\textbf{T}}}^{\left( D\right) }\left( \beta _{1};\beta
_{2}\right) =\frac{1}{\mathcal{N}\pi ^{2}}\times  \notag \\
&&\int d^{2}\beta _{3}d^{2}\beta _{4}\chi _{1234}\left( \beta _{1};\beta
_{2};\beta _{3};\beta _{4}\right)  \notag \\
&&\times \chi _{3}^{\left( D\right) }\left( \beta _{3}\right) \chi
_{4}^{\left( D\right) }\left( \beta _{4}\right) \; ,
\label{chi12_ideal}
\end{eqnarray}
where $d^{2}\beta _{k}=d\beta _{k}d\beta _{k}^{\ast }$ with $\beta _{k}$
complex coherent amplitude, and
$\chi _{1234}\left( \beta _{1};\beta _{2};\beta _{3};\beta _{4}\right)$ is
the characteristic function of the initial state. It corresponds to
$|\Phi>_{1234}$, see Eq.~(\ref{phi_BS}), for the ideal scheme, and to
$|\Phi^{\prime }>_{1234}$, see Eq.~(\ref{phi_loss}), for the realistic scheme.
In the above, $\chi _{k}^{\left( D\right) }\left( \beta _{k}\right) $ denotes
the characteristic function of the conditional measurement realized by
detectors $D_{3}$ and $D_{4}$ on the modes $k=3,4$. For more details, see
Appendix A, in which $D=\left\vert 1\right\rangle \left\langle 1\right\vert $ when
postselection is applied using single-photon projectors and $D=on$ when
postselection is applied using realistic on-off operators (POVMs).

We will consider the following states:

\begin{itemize}
\item \textit{Theoretical states} -- These are the ideal states defined in
Table~\ref{TheorDef}. They are not always exactly attainable within our
scheme of generation, not even in ideal conditions. Their performances as
entangled resources have been investigated in~\cite{Dellanno2007,Dellanno2010}.

\item \textit{States generated experimentally: ideal conditions} -- These are
the states generated by our scheme when we assume that losses are absent,
detectors are perfectly photon-resolving, and measurements are perfectly projective.

\item \textit{States generated experimentally: realistic conditions} -- These
are the states generated by our scheme when losses are considered, and only
on/off measurements, described by non--ideal POVMs, are allowed.
\end{itemize}

In the formalism of the characteristic function, the fidelity of teleportation
is defined as
\begin{equation}
\mathcal{F}=\frac{1}{\pi }\int d^{2}\lambda \chi _{in}(\lambda )
\chi_{out}(-\lambda ) \; ,
\label{Fidelity}
\end{equation}
where $d^{2}\lambda =d\lambda d\lambda^{\ast }$with $\lambda $ the vector
of complex coherent amplitude for a generic state. For an input coherent
state $|\alpha >$, the characteristic function $\chi _{in}\equiv \chi _{coh}$
is
\begin{equation}
\chi _{coh}(\lambda )=e^{-\frac{1}{2}|\lambda |^{2}+2i \Im \left[ \lambda
\alpha ^{\ast }\right] } \; ,
\label{coh_state}
\end{equation}
while the characteristic function $\chi _{out}$ of the output state
is~\cite{Dellanno2007}
\begin{equation}
\chi _{out}(\lambda )=\chi _{coh}(\lambda )\chi _{res}(\lambda ^{\ast}; \lambda) \; ,
\label{chi_out}
\end{equation}
where $\chi _{res}\left( \lambda ^{\ast };\lambda \right)$ denotes the
characteristic function of the entangled state used as resource for the
protocol. Before proceeding further, we recall that the generation scheme is
based on the condition $\kappa ^{2}<<1$ and on the possibility of optimizing
over largely tunable parameters. The only unconditioned parameter is the
amplitude $r$ of the squeezing operator $S_{12}(\xi )$. Once $r$ is fixed,
the fidelity of the state generated in the scheme will depend on the two
squeezing parameters and on transmissivities, therefore, from now on, we
will redefine the fidelity $\mathcal{F}$ as:
$\mathcal{F}_{\text{\textbf{T}}}(\zeta ,\xi ,T)$.

This notation allows to see clearly that the optimization has to be
performed with respect to the phases $\phi _{\zeta },\phi _{\xi }$ of the
two squeezing operators, the transmissivity $T$ (recall that $T_{1}=T_{2}=T$),
and the squeezing amplitude $s$ of the ancillary squeezing operator
$S_{34}(\xi )$. In the following we will show that optimization with respect
to phases and transmissivities is compatible with the assumptions imposed in
order to generate experimentally the class of squeezed Bell states. In
general, at fixed squeezing amplitude $|\zeta |=r$ for modes $1$ and $2$,
see Fig.~(\ref{FigSchemeIdeal}), the optimal fidelity is defined as
\begin{equation}
\mathcal{F}_{opt}(r)=\underset{\phi _{\zeta },\xi ,T}{\max }
\mathcal{F}_{\text{\textbf{T}}}(\zeta ,\xi ,T) \; .
\label{maxFid}
\end{equation}
Starting from this general relation, one has to solve the optimization
problem with respect to the phases of the complex squeezing amplitudes.
A thorough analysis shows that the optimization procedure always yields
$\phi _{\zeta }=\phi _{\xi }=\pi $, thus implying that the optimal
building blocks of the generation scheme, see Fig.~(\ref{FigSchemeIdeal}),
are always two independent two--mode squeezed states with $\zeta =-r$,
and $\xi =-s$. This finding is in agreement with and justifies \emph{a posteriori}
the \emph{a priori} position assumed in the previous section. Therefore,
from now on we will fix for the squeezing operators the notations:
$S_{12}(-r),S_{34}(-s)$.

The optimization with respect to $T$ must take into account the role that
the transmissivity plays in setting the distillation success rate, see
Eq.~(\ref{N_T_on}). Furthermore, the result of this analysis must be congruent
with the assumption $\kappa^2 << 1$, implying the \emph{high transmissivity}
$T = \cos^2 |k|$ needed to implement the generation scheme.

The fidelity turns out to be a monotonically increasing function of $T$. The
optimal value is thus obtained asymptotically for $T$ approaching unity.
Since the limiting value $T=1$ corresponds, obviously, to a vanishing
success rate, in the following we will set $T=0.99$ (a value that is
perfectly reachable in a real experiment) and remove the dependence on the
transmissivity. In this way, one satisfies the assumption $\kappa^2 \sim
0.01 << 1$ and, simultaneously, achieves \emph{de facto} the optimization
with respect to the transmissivity.

Finally, one is left with the optimization with respect to the ancillary
squeezing parameter $s$. For each case one can identify the explicit value
of $s$ that, at each given $r$, maximizes the fidelity. We will see that for
very small values of $r$ the optimization selects non-Gaussian states that
coincide essentially with the de-Gaussified photon-subtracted squeezed
states. In a regime of intermediated values of $r$ the optimization selects
non-Gaussian states that coincide essentially with the squeezed Bell states.
Finally, for large values of $r$ all states converge to the
continuous-variable Einstein-Podolsky-Rosen state and therefore the
performance of Gaussian twin beams is indistinguishable from that of
non-Gaussian squeezed Bell states.

\subsection{Ideal single-photon measurements}

To begin with, let us consider the teleportation fidelity in the ideal case
where the detectors $D_3$ and $D_4$, see Fig.~(\ref{FigSchemeIdeal}), realize
simultaneous projective single-photon measurements, and there are no losses.
Under these conditions, the output state is pure and of the form given by
Eq.~(\ref{psi_12}).

\begin{figure}[ht]
\includegraphics*[width=8cm]{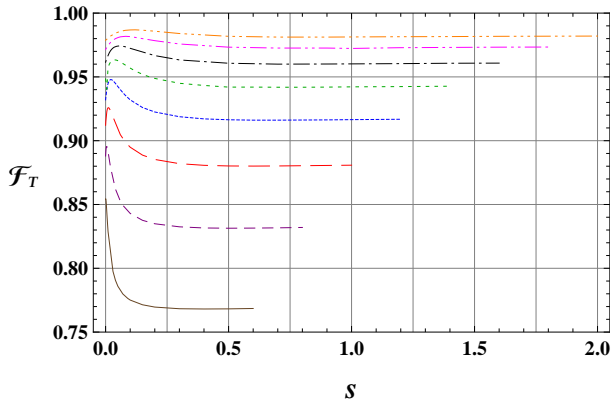}
\caption{Fidelity of teleportation, as a function of
the ancillary squeezing parameter $s$ ($\leq r$), obtained
using as shared entangled resources, in the Braunstein-Kimble-Vaidman
teleportation protocol, the tunable states generated
by our ideal scheme (perfect single-photon conditional measurements).
The fidelity is plotted parametrically for different values of the main squeezing
parameter $r$: (a) $r=0.6$ (brown full line); (b) $r=0.8$ (purple dashed line); (c) $r=1$ (red large--dashed
line); (d) $r=1.2$ (blue dotted line); (e) $r=1.4$ (green large--dotted
line); (f) $r=1.6$ (black dotted--dashed line); (g) $r=1.8$ (magenta double
dotted--dashed line); (h) $r=2$ (orange triple dotted--dashed line). The
point at $s=0$ corresponds to the fidelity achieved with a photon-subtracted
squeezed state generated in ideal conditions, while at $s=r$ one recovers the
fidelity achieved with an ideal twin beam.}
\label{FigFidProjVsS}
\end{figure}

In Fig.~(\ref{FigFidProjVsS}) we analyze the fidelity
$\mathcal{F}_{\text{\textbf{T}}}$ for the teleportation of an unknown single-mode coherent state
using, as shared entangled resource, the ideal states of Eq.~(\ref{psi_12}).
The teleportation fidelity is plotted as a function of the ancillary
squeezing $s$, for all values of $s \leq r$ and for eight different values
of $r$. For each curve, the entangled resource corresponding to $s=0$ is a
de-Gaussified photon-subtracted squeezed state, while for $s=r$ the
corresponding entangled resource is a Gaussian twin beam.

It can be seen that the maximum of the teleportation fidelity moves toward
higher values of $s$ as $r$ increases; at the same time the maximum becomes
less pronounced. The results can be summarized as follows:

\begin{itemize}
\item For small values of the main squeezing $r$ the optimal resource for teleportation
is obtained for a vanishingly small ancillary squeezing $s$. In particular, for the sequence of
values $r=0.6,0.8,1$ the approximate squeezed Bell state, as realized by our generation scheme,
yields the best performance, and the ancillary squeezing $s$ does not exceed, in order of
magnitude, $\kappa ^{2}\sim 0.01$ (see Table~\ref{sVSk}).

\item For values of $r$ greater than $1$ the state produced by the
generation scheme and corresponding to the maximum fidelity, as $r$ grows
moves away increasingly from the squeezed Bell state (the value of $s$
exceeds sensibly the order of magnitude of $\kappa ^{2}\sim 0.01$, see
Table~\ref{sVSk}). However, this state still provides a better performance than that of a
twin beam and of an (experimentally generated) photon-subtracted squeezed
state.

\item In this same last region a Gaussian twin beam provides a better performance
than that of the photon-subtracted squeezed states and approximate squeezed Bell
states generated by our scheme in ideal conditions.
\end{itemize}

\begin{table}[tp]
\begin{tabular}{|c|c|}
\hline
$r$ & \multicolumn{1}{||c|}{$s$} \\ \hline\hline
0.6 & 0.00057 \\ \hline
0.8 & 0.0046 \\ \hline
1.0 & 0.011 \\ \hline
1.2 & 0.022 \\ \hline
1.4 & 0.036 \\ \hline
1.6 & 0.056 \\ \hline
1.8 & 0.082 \\ \hline
2.0 & 0.12 \\ \hline
\end{tabular}
\caption{Values of the ancillary squeezing $s$ corresponding to the maximum performance
of the states produced by our scheme in the ideal case for the given values of
the principal squeezing $r$.}
\label{sVSk}
\end{table}

\begin{figure}[ht]
\includegraphics*[width=8cm]{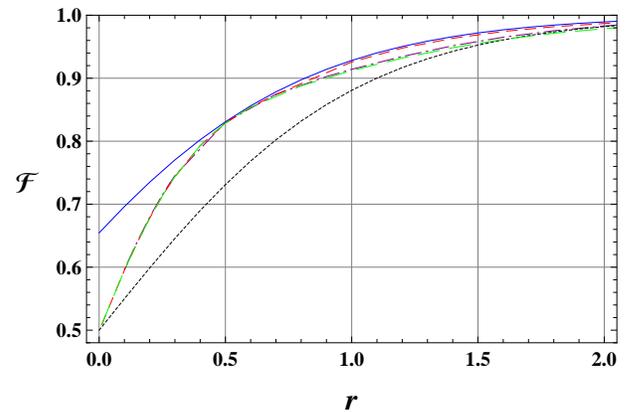}
\caption{Comparison among the optimized fidelity of teleportation, Eq.~(\ref{maxFid}),
obtained using as entangled resources the class of optimized tunable states produced by our scheme in ideal conditions (red dashed line), the fidelity of teleportation obtained using as entangled resources the photon-subtracted squeezed state generated in ideal conditions (green large--dashed line), the ideal optimized fidelity obtained using as entangled resources the theoretical squeezed Bell states (cyan solid line), the fidelity obtained using as entangled resources the theoretical photon-subtracted squeezed states (purple dotted--dashed line), and the fidelity obtained using as entangled resources the theoretical
Gaussian twin beams (black dotted line).}
\label{FigProjIdeal}
\end{figure}

We will now compare the optimal fidelity of teleportation, Eq.~(\ref{maxFid}, that can be achieved using as entangled resources the states in the class produced by our scheme, \textit{i.e.} the value of the
maximum in Fig.~(\ref{FigFidProjVsS}), with the ideal fidelity of teleportation obtained when the
entangled resources are the theoretical states listed in Table~\ref{TheorDef}.
In Fig.~(\ref{FigProjIdeal}) we report, as a function of the principal squeezing $r$,
the behavior of the optimal fidelity corresponding to the resource states generated
by our ideal scheme, and we compare it with that associated to the theoretical states
listed in Table~\ref{TheorDef} (twin beams, photon-subtracted squeezed states, and
squeezed Bell states). In the same figure we report also the fidelity of
the photon-subtracted squeezed states ($s=0$) generated by the ideal scheme.

From this analysis, it emerges that in the ideal contest of state generation, the ideal optimized
fidelity of teleportation is achieved, in the entire range of the considered values of $r$,
by using as entangled resources the (optimized) theoretical squeezed Bell
states~\cite{Dellanno2007}. On the other hand, the optimal fidelity achievable using
the class of states that can be generated by our experimental scheme in ideal conditions
approximates remarkably well the ideal one associated to the theoretical squeezed Bell states
for large enough values of the principal squeezing $r$.

\begin{figure}[ht]
\includegraphics*[width=8cm]{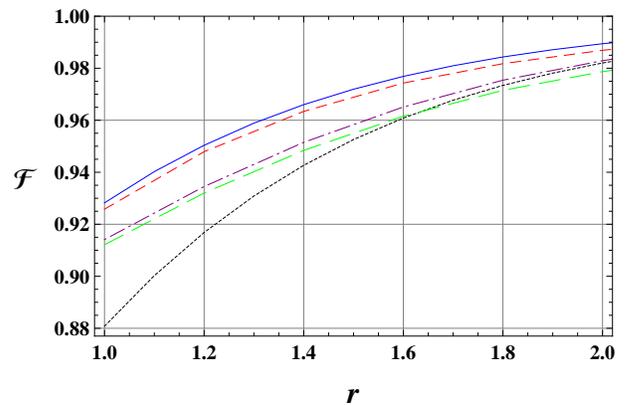}
\caption{Zoom of Fig.~(\ref{FigProjIdeal}) in the
range $1\leq r\leq 2$ for: the optimized fidelity of teleportation obtained
considering as entangled resources the optimized tunable states generated by our
scheme in ideal conditions (red dashed line); the fidelity of teleportation obtained
considering as entangled resources the photon-subtracted squeezed states
generated in ideal conditions (green large--dashed line); the optimized fidelity
of teleportation obtained considering as entangled resources the theoretical squeezed Bell
states (cyan solid line); the fidelity of teleportation obtained considering as entangled
resources the theoretical photon-subtracted squeezed states (purple dotted--dashed line);
and the fidelity of teleportation obtained considering as entangled resources the theoretical
Gaussian twin beams (black dotted line).}
\label{FigProjIdealPart1}
\end{figure}

It is also important to notice that while the fidelities associated to the theoretical states
and to the photon-subtracted squeezed states generated in ideal conditions can be
computed analytically as functions of $r$, the optimal fidelities associated to the entire class
of states produced by our scheme in ideal conditions must be determined numerically point by point,
so that the plots of these optimized fidelities, if seen in greater detail, would look as discrete,
broken lines. In the plot range\textit{\ }$0<$\textit{\ }$r\lesssim 2$ that represents the current
levels of squeezing that are experimentally feasible~\cite{Schnabel1}, we can then identify two
distinct regimes:

a) $r\lesssim 0.5$ \textit{--} the procedure of maximization Eq.~(\ref{maxFid})
yields $s\simeq 0$, \textit{i.e.} the best entangled resources generated
by our scheme in ideal conditions coincide with states that approximate the
photon-subtracted squeezed states generated in ideal conditions. On the other hand, the three
curves corresponding, respectively, to the the optimized fidelity of teleportation, Eq.~(\ref{maxFid},
obtained using as entangled resources the class of states produced by our scheme in ideal
conditions, to the fidelity of teleportation obtained using as entangled resources the photon-subtracted squeezed state generated in ideal conditions, and to the fidelity of teleportation obtained using as entangled resources the theoretical photon-subtracted squeezed states, are superimposed and lie in between
an upper limit given by the fidelity of teleportation obtained using as entangled resources the optimized theoretical squeezed Bell states and a lower limit given by the fidelity of teleportation obtained using as entangled resources the theoretical Gaussian twin beams.

b) $r>0.5$ \textit{--} the optimized resources generated
by our scheme outperform both the theoretical photon-subtracted squeezed states
and those generated in ideal conditions, at the same time providing a performance
very close to that of the optimized theoretical squeezed Bell states.
In Fig.~(\ref{FigProjIdealPart1}) we report their behaviors in the range $1\leq r\leq 2$.
As an example, if we fix $r=1.6$, we obtain the value $0.974$ (at $s=0.056$)
for the optimized fidelity of teleportation, Eq.~(\ref{maxFid}, obtained using as entangled resources the photon-subtracted squeezed state generated in ideal conditions. At the same
value of $r$, the teleportation fidelities obtained using as entangled resources
the theoretical states are, respectively, $0.977$ for the optimized theoretical
squeezed Bell state; $0.965$ for the theoretical photon-subtracted squeezed state; and $0.961$
for the theoretical Gaussian twin beam. Therefore, within the ideal conditions considered so far,
the level of performance of the states generated by our scheme as entangled resources for quantum teleportation is remarkably close to that of the ideal theoretical states.

\subsection{Realistic conditions}

A realistic scenario of state generation within our scheme must include inefficient photon
detection and a lossy environment for the input pair of two--mode
squeezed states. In what follows we have considered the value $\eta =0.15$
for the detection efficiency (that is the value currently obtainable in real
experiments). Moreover, we remark that the values of the squeezing amplitude
$r$ which appear in the plots are referred to the theoretical principal squeezing,
but the reduction to the effective squeezing $r^{^{\prime }}$ has been taken
into account when displaying the final results.

\begin{figure}[ht]
\includegraphics*[width=8cm]{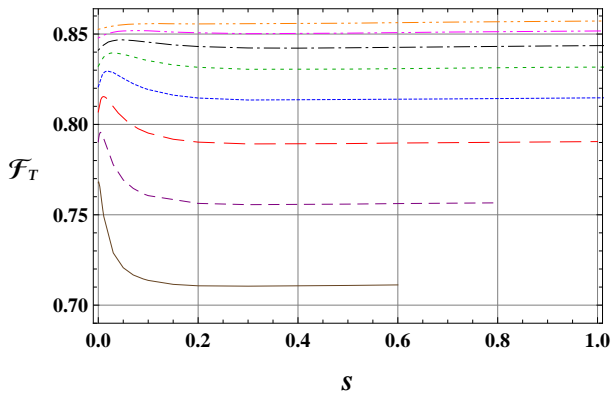}
\caption{Fidelity of teleportation in a realistic lossy
scenario (level of losses equal to $0.15$, \textit{i. e.} $T_{\ell }=0.85$, and
$\protect\eta =0.15$), as a function of the ancillary squeezing $s$ ($\leq r$) for
the same values of the principal squeezing $r$ of Fig.~(\ref{FigFidProjVsS}):
(a) $r=0.6$ (brown solid line); (b) $r=0.8$ (purple dashed line); (c) $r=1$
(red large--dashed line); (d) $r=1.2$ (blue dotted line); (e) $r=1.4$ (green large--dotted line);
(f) $r=1.6$ (black dotted--dashed line); (g) $r=1.8$ (magenta double dotted--dashed line);
(h) $r=2$ (orange triple dotted--dashed line).}
\label{FigFidPovmRealVsS}
\end{figure}

In Fig.~(\ref{FigFidPovmRealVsS}) we have plotted the optimized fidelity of
teleportation, that depends on the squeezing amplitudes $r$ and $s$, assuming
an overall transmissivity $T_{\ell }=0.85$, \textit{i. e.} a level of loss equal
to $0.15$, in Eq.~(\ref{phi_loss}). For consistency in the comparison between
ideal and realistic conditions, in the figure we have plotted the optimized
fidelity as a function of the ancillary squeezing $s$ ($\leq r$), assuming
for the principal squeezing $r$ the same values as in Fig.~(\ref{FigFidProjVsS}).
We can observe that:

a) The overall behavior of the fidelities does not change qualitatively, apart from
a smoothing of the curves around their maximum.

b) As expected, the fidelities suffer a further deterioration due to the
combined effect of losses and non--ideal single photon detection processes.

In this first plot the level of losses equals $0.15$. At present, this level
is experimentally accessible by properly choosing optical components for the
source of squeezing. On the other hand, very recently an outstanding source
of squeezing with an overall loss of less than $0.08$ has been reported~\cite{Schnabel2}.
In view of this result, we have considered the behavior of the fidelities
when the level of losses is varied.

\begin{figure}[ht]
\includegraphics*[width=8cm]{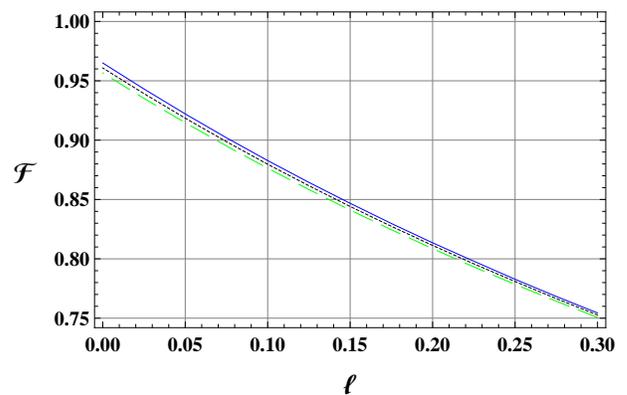}
\caption{Optimized fidelity of teleportation (blue solid line) obtained
considering as entangled resources the optimized tunable states generated by our
scheme in realistic conditions, with $\eta =0.15$, plotted as a function of
the loss parameter $\ell $, at fixed principal squeezing $r=1.6$. The optimized fidelity
is compared with those obtained considering as entangled resources the photon-subtracted
squeezed states ($s=0$, black dotted line)) and the Gaussian twin
beams ($s=r$, green large--dashed line), generated in the same realistic
conditions.}
\label{FigFidLosses}
\end{figure}

Fixing the detection efficiency at $\eta =0.15$ and the principal squeezing parameter at $r=1.6$, in Fig.~(\ref{FigFidLosses}) we report the optimized fidelity of teleportation, Eq.~(\ref{maxFid}, obtained
considering as entangled resources the optimized tunable states generated by our scheme in realistic conditions, as a function of the loss parameter, denoted by $\ell $. In the same plot the optimal fidelity
is compared with the fidelity of teleportation obtained considering as entangled resources the photon-subtracted squeezed states and the Gaussian twin beams generated in the same realistic conditions,
\textit{i.e.} with $\eta =0.15$ and $r=1.6$. As it can be seen, for losses up to $\ell =0.30$ the
optimized tunable states generated by our scheme in realistic conditions always yield the largest fidelity
of teleportation. It has to be noted that, at fixed principal squeezing $r$, the value of the ancillary squeezing $s$ corresponding to the maximum value of the optimized fidelity remains essentially constant. Indeed, in the case considered, this value varies in the interval $[0.048,0.050]$.

From Fig.~(\ref{FigFidLosses}) we see that in realistic conditions and for values of the principal squeezing $r$ varying in the interval $[1.2,1.6]$ the optimized tunable states yield
fidelities of teleportation sizeably larger than those provided by
photon-subtracted squeezed states and Gaussian twin beams.
Furthermore, the behavior reported in Fig.~(\ref{FigFidLosses}) implies that
foreseeable improvements in the control of losses could lead to levels of
performance of the tunable non-Gaussian states comparable to those of the
theoretical squeezed Bell states.

\section{Discussion and outlook}

In the present work we have introduced a scheme of state generation able to
produce a class of tunable two-mode non-Gaussian states that approximate closely the
class of theoretical squeezed Bell states introduced in Refs.~\cite{Dellanno2007,Dellanno2010}.
A thorough analysis yields that the states generated by our scheme in realistic conditions,
when properly optimized by tuning two experimentally adjustable free parameters,
provide, as entangled resources, the maximum fidelity of teleportation in
the Braunstein-Kimble-Vaidman teleportation protocol of an unknown coherent state.
Indeed, the optimized tunable non-Gaussian resources yield, in the most interesting
range of the currently accessible experimental values of the principal squeezing amplitude $r$,
a better performance both with respect to Gaussian twin beams and to
photon-subtracted squeezed states, the latter being at present the best performing
continuous-variable entangled resource that can be produced experimentally.
This result holds true both in ideal and in realistic conditions. In particular, in ideal conditions
of generation (no losses, perfect photon-resolving detection, perfect projections), for values of
the principal squeezing $r>0.5$, the optimized tunable states show a level of performance very close
to that of the optimized theoretical squeezed Bell states. In realistic conditions (presence
of losses, only on/off measurements allowed), the optimized tunable states provide
again, in a wide interval of values of the principal squeezing $r$, the best performance
with respect to that yielded by Gaussian twin beams state and photon-subtracted squeezed states.

It is interesting to note that even a slight improvement, with respect to the current
experimental situation, in reducing the level of losses and in increasing the detection
efficiency would lead to a significant improvement in the performance of the optimal
tunable states generated by our scheme in realistic conditions. As remarked in subsection III.B,
a sizeable reduction of losses to very low levels seems at hand. Regarding the problem of improving the efficiency in photon-resolving procedures, detectors based on superconducting devices
could lead to important progress in the near future~\cite{Hadfield}.

The theoretical study carried out in the present work proves
that our scheme of state generation can produce a large class of two-mode non-Gaussian states
that, when operated as entangled resources, can outperform both the currently available
entangled two-mode Gaussian and de-Gaussified states. In forthcoming works, the
experimental set up needed to realize our scheme of generation will be
designed, analyzed, and discussed in all its technical details. To this end, we will consider two possible
working regimes: continuous-wave regime and pulsed regime. We will
also consider at length the problem of efficient detection in coincidence of two photons in two different modes, as this is one of the crucial requirements of our scheme, and we will show how the generated states can be reconstructed by performing suitable tomographic homodyne detections. Finally, further aspects
of tunable non-Gaussian states will be investigated, in particular concerning protocols of entanglement
swapping and distillation, as well as their properties with respect to Bell's nonlocality and
Einstein-Podolsky-Rosen steering.

\section{Acknowledgement}

Daniela Buono, Fabio Dell'Anno, Silvio De Siena, and Fabrizio Illuminati acknowledge support from the EU STREP Project iQIT, Grant Agreement No. 270843.

\section{Appendix A}

\vspace{0.5cm}

\noindent \emph{Postselection: single-photon projector --}

In the ideal scheme of state generation, Fig.~(\ref{FigSchemeIdeal}), the post-selection
strategy is based on ideal conditional measurements, \textit{i.e.} simultaneous
detections of single photons in mode $3$ and $4$. Such coincidence
detections of single photons project the density matrix $\rho _{1234}$ into
the \textit{tunable} state $\rho _{\text{\textbf{T}}}^{\left( |1><1|\right)}$,
which reads
\begin{align*}
& \rho _{\text{\textbf{T}}}^{\left( |1><1|\right) } \\
& \equiv \frac{\text{Tr}_{34}\left[ \rho _{1234}\otimes \text{ }_{3}|1><1|_{3} \otimes
\text{ }_{4}|1><1|_{4}\right] }{\mathcal{N}_{\text{\textbf{T}}}^{\left( |1><1|\right) }} \\
& =\frac{1}{\mathcal{N}_{\text{\textbf{T}}}^{\left( |1><1|\right) }\pi^{4}}
\int d^{2}\beta _{1}d^{2}\beta _{2}d^{2}\beta _{3}d^{2}\beta _{4} \\
& \times \chi _{1234}\left( \beta _{1};\beta _{2};\beta _{3};\beta
_{4}\right) D_{1}\left( -\beta _{1}\right) D_{2}\left( -\beta _{2}\right) \\
& \times \text{Tr}_{\text{ }34}\left[ D_{3}\left( -\beta _{3}\right)
D_{4}\left( -\beta _{4}\right) _{3}|1><1|_{3\text{ }4}|1><1|_{4}\right]
\text{ } \\
& =\frac{\int d^{2}\beta _{1}d^{2}\beta _{2}\mathcal{M}^{\left(
|1><1|\right) }\left( \beta _{1};\beta _{2}\right) D_{1}\left( -\beta
_{1}\right) D_{2}\left( -\beta _{2}\right) }{\mathcal{N}_{\text{\textbf{T}}}^{\left(
|1><1|\right) }\pi ^{2}} \, .
\end{align*}

In the above,

\begin{align*}
& \mathcal{M}^{\left( |1><1|\right) }\left( \beta _{1};\beta _{2}\right) \\
& \equiv \frac{1}{\pi ^{2}}\int d^{2}\beta _{3}d^{2}\beta _{4}\chi
_{1234}\left( \beta _{1};\beta _{2};\beta _{3};\beta _{4}\right) \\
& \times \chi _{3}^{\left( |1>\right) }\left( \beta _{3}\right) \chi
_{4}^{\left( |1>\right) }\left( \beta _{4}\right) \, .
\end{align*}
Recalling that
$$
<m|D\left( -\alpha \right) |n>=\left( \frac{n!}{m!}\right)
^{1/2}\alpha ^{m-n}e^{-|\alpha |^{2}/2}L_{n}^{m-n}\left( |\alpha
|^{2}\right) \; ,
$$
and that $L_{1}^{0}\left( x\right) =L_{1}\left( x\right) =1-x$, one has
\begin{align*}
\chi _{k}^{\left( |1><1|\right) }\left( \beta _{k}\right) & =\text{Tr}_{k}
\left[ \hat{D}_{l}\left( -\beta _{l}\right) _{3}|1><1|_{3}\right] \\
& =\left( 1-|\beta _{k}|^{2}\right) e^{-|\beta _{k}|^{2}/2}\text{ \ \ for }
k=3,4\; ,
\end{align*}
which is the characteristic function of the single-photon projector,
$_{k}|1><1|_{k}$, acting on the $k-$th mode. Moreover, the normalization constant
$\mathcal{N}_{\text{\textbf{T}}}^{\left( |1><1|\right) }$ is given by
\begin{equation*}
\mathcal{N}_{\text{\textbf{T}}}^{\left( \left\vert 1\right\rangle
\left\langle 1\right\vert \right) }\mathcal{=}\text{Tr}_{1234}\left[ \rho
_{1234}\otimes \text{ }_{3}\left\vert 1\right\rangle \left\langle
1\right\vert _{3}\otimes \text{ }_{4}\left\vert 1\right\rangle \left\langle
1\right\vert _{4}\right] \; .
\end{equation*}
In conclusion, we have
\begin{align*}
& \chi _{\text{\textbf{T}}}^{\left( \left\vert 1\right\rangle \left\langle
1\right\vert \right) }\left( \gamma _{1},\gamma _{2}\right) \\
& =\frac{\text{Tr}_{12}\left[ \rho _{12}D_{1}\left( \gamma _{1}\right)
D_{2}\left( \gamma _{2}\right) \right] }{\mathcal{N}_{\text{\textbf{T}}}^{\left( \left\vert 1\right\rangle \left\langle 1\right\vert \right) }} \\
& =\frac{1}{\mathcal{N}_{\text{\textbf{T}}}^{\left( \left\vert
1\right\rangle \left\langle 1\right\vert \right) }\pi ^{2}}\int d^{2}\beta
_{1}d^{2}\beta _{2}\mathcal{M}^{\left( |1><1|\right) }\left( \beta
_{1};\beta _{2}\right) \\
& \times \text{Tr}_{12}\left[ D_{1}\left( \gamma _{1}\right) D_{1}\left(
-\beta _{1}\right) D_{2}\left( \gamma _{2}\right) D_{2}\left( -\beta
_{2}\right) \right] \\
& =\frac{1}{\mathcal{N}_{\text{\textbf{T}}}^{\left( \left\vert
1\right\rangle \left\langle 1\right\vert \right) }}\mathcal{M}^{\left(
|1><1|\right) }\left( \gamma _{1};\gamma _{2}\right) \; .
\end{align*}

\vspace{0.5cm}

\noindent \emph{Postselection: realistic on/off operator (POVM)--}

In the realistic case, see Fig.~(\ref{SchemeReal}), the post-selection
strategy is based on realistic conditional measurements made by realistic
on/off detectors. They are described by positive operator-valued measures
(POVMs), Eq.~(\ref{POVMid}). The detection $on-$POVM yields the state

\begin{align*}
\rho _{\text{\textbf{T}}}^{\left( on\right) }& =\frac{\text{Tr}_{34}\left[
\rho _{1234}\otimes \Pi _{3}^{(on)}\left( \eta _{3}\right) \otimes \Pi
_{4}^{(on)}\left( \eta _{4}\right) \right] }{\mathcal{N}_{\text{\textbf{T}}}^{\left( on\right) }} \\
& =\frac{1}{\mathcal{N}_{\text{\textbf{T}}}^{\left( on\right) }\pi ^{4}}\int
d^{2}\beta _{1}d^{2}\beta _{2}d^{2}\beta _{3}d^{2}\beta _{4} \\
& \times \chi _{1234}\left( \beta _{1};\beta _{2};\beta _{3};\beta
_{4}\right) D_{1}\left( -\beta _{1}\right) D_{2}\left( -\beta _{2}\right) \\
& \times \text{Tr}_{34}\left[ D_{3}\left( -\beta _{3}\right) D_{4}\left( -\beta _{4}\right)
\Pi _{3}^{(on)}\left( \eta _{3}\right) \Pi_{4}^{(on)}\left( \eta _{4}\right) \right] \\
& =\frac{1}{\mathcal{N}_{\text{\textbf{T}}}^{\left( on\right) }\pi^{4}}
\int d^{2}\beta _{1}d^{2}\beta _{2}\mathcal{M}^{\left( on\right) }\left( \beta_{1};\beta _{2}\right) \\
& \times D_{1}\left( -\beta _{1}\right) D_{2}\left( -\beta _{2}\right) \; ,
\end{align*}
where
\begin{align*}
& \mathcal{M}^{\left( on\right) }\left( \beta _{1};\beta _{2}\right) \\
& =\frac{1}{\mathcal{N}_{\text{\textbf{T}}}^{\left( on\right) }\pi ^{2}}\int
d^{2}\beta _{3}d^{2}\beta _{4}\chi _{1234}\left( \beta _{1};\beta _{2};\beta_{3};\beta _{4}\right) \\
& \times \chi _{3}^{on}\left( \beta _{3}\right) \chi _{4}^{on}\left( \beta_{4}\right) \; .
\end{align*}
Here
\begin{align*}
\chi _{k}^{on}\left( -\beta _{l}\right) & \equiv \text{Tr}_{k}\left[ D_{k}\left( -\beta _{k}\right)
\Pi _{k}^{(on)}\right] \\
& =\pi \delta ^{(2)}\left( \beta _{k}\right) -\frac{1}{\eta _{k}}\exp
\left\{ -\frac{2-\eta _{k}}{2\eta _{k}}|\beta _{k}|^{2}\right\} \\
& =\chi _{k}^{on}\left( \beta _{k}\right)
\end{align*}
is the characteristic function of the POVM of the photo-detector of the modes
$3$ and $4$, and the normalization reads
\begin{equation*}
\mathcal{N}_{\text{\textbf{T}}}^{\left( on\right) }=\text{Tr}_{1234}\left[
\rho _{1234}\otimes \Pi _{3}^{(on)}\left( \eta _{3}\right) \otimes \Pi
_{4}^{(on)}\left( \eta _{4}\right) \right] \; .
\end{equation*}

The characteristic function corresponding to the density matrix
$\rho _{\text{\textbf{T}}}^{\left( on\right) }$ is
\begin{align*}
\chi _{\text{\textbf{T}}}^{(on)}\left( \gamma _{1};\gamma _{2}\right) & =
\frac{1}{\mathcal{N}_{\text{\textbf{T}}}^{\left( on\right) }}\text{Tr}_{12}
\left[ \rho _{12}D_{1}\left( \gamma _{1}\right) D_{2}\left( \gamma_{2}\right) \right] \\
& =\frac{1}{\mathcal{N}_{\text{\textbf{T}}}^{\left( on\right) }\pi ^{2}}\int
d^{2}\beta _{1}d^{2}\beta _{2}\mathcal{M}^{\left( on\right) }\left( \beta_{1};\beta _{2}\right) \\
& \times \text{Tr}_{12}\left[ D_{1}\left( -\beta _{1}\right) D_{2}\left(
-\beta _{2}\right) D_{1}\left( \gamma _{1}\right) D_{2}\left( \gamma_{2}\right) \right] \\
& =\frac{1}{\mathcal{N}_{\text{\textbf{T}}}^{\left( on\right) }\pi^{2}}
\mathcal{M}^{\left( on\right) }\left( \gamma _{1};\gamma _{2}\right) \; .
\end{align*}

In terms of the complex amplitudes $\beta _{1}$ and $\beta _{2}$,
the characteristic function reads
\begin{align*}
& \chi _{\text{\textbf{T}}}^{(on)}\left( \beta _{1};\beta _{2}\right) \\
& =\frac{1}{\mathcal{N}_{\text{\textbf{T}}}^{\left( on\right) }\pi ^{2}}
\int d^{2}\beta _{3}d^{2}\beta _{4}\chi _{1234}\left( \beta _{1};\beta _{2};
\beta_{3};\beta _{4}\right) \\
& \times \chi _{3}^{on}\left( \beta _{3}\right) \chi _{4}^{on}\left( \beta_{4}\right) \\
& =\frac{1}{\mathcal{N}_{\text{\textbf{T}}}^{\left( on\right) }}\left[ \chi_{1234}
\left( \beta _{1};\beta _{2};0;0\right) \right. \\
& +\frac{1}{\pi }\int d^{2}\beta _{4}\chi _{1234}\left( \beta _{1};\beta
_{2};0;\beta _{4}\right) \mathcal{G}_{4}\left( \beta _{4}\right) \\
& +\frac{1}{\pi }\int d^{2}\beta _{3}\chi _{1234}\left( \beta _{1};\beta
_{2};\beta _{3};0\right) \mathcal{G}_{3}\left( \beta _{3}\right) \\
& \left. +\frac{1}{\pi ^{2}}\int d^{2}\beta _{3}d^{2}\beta _{4}\chi
_{1234}\left( \beta _{1};\beta _{2};\beta _{3};\beta _{4}\right) \mathcal{G}
_{3}\left( \beta _{3}\right) \mathcal{G}_{4}\left( \beta _{4}\right) \right]
\; ,
\end{align*}
where
\begin{eqnarray*}
\mathcal{G}_{k}\left( \beta _{k}\right) &=&-\frac{1}{\eta _{k}}\exp \left\{ -
\frac{2-\eta _{k}}{2\eta _{k}}|\beta _{k}|^{2}\right\} \; . \\
&&\left( k=3,4\right) \; .
\end{eqnarray*}

\vspace{0.5cm}

\emph{Effective values of the squeezing amplitudes --}

The values of the principal squeezing $r$ and of the ancillary squeezing $s$
are referred to the pure input parameters, before decoherence and losses
affect the incoming beams and decrease the amplitudes to the real
values $r^{\prime }$ and $s^{\prime }$.
In the following we determine the relation holding between the squeezing parameters before
and after the action of decoherence and losses. For this purpose, we express
the two-mode squeezing operator $S_{ab}\left( -|\lambda |\right) $ in terms
of single-mode squeezing operators $S_{c}\left( -|\lambda |\right)
,S_{d}\left( |\lambda |\right) $. These are obtained by introducing the
transformed annihilation operators $c$ and $d$ defined by the linear
superpositions
\begin{eqnarray*}
c &=&\frac{a+b}{\sqrt{2}} \; , \\
d &=&\frac{-a+b}{\sqrt{2}} \; .
\end{eqnarray*}
Under this transformation, the two-mode squeezed state $S_{ab}\left( -|\lambda
|\right) |\mathbf{0}>_{ab}$ goes into the two-mode squeezed state
\begin{equation}
S_{c}\left( -|\lambda |\right) S_{d}\left( |\lambda |\right) |\mathbf{0}
>_{cd} \; ,  \label{cd_modes}
\end{equation}
where $S_{k}\left( \lambda \right) =\exp \left[ -\frac{1}{2}\lambda k^{\dag
2}+\frac{1}{2}\lambda ^{\ast }k^{2}\right] ,\left( k=c,d\right) $ denotes the
single-mode squeezing operator. Introducing the fictitious beam splitters
that simulate losses and decoherence, the original, decoherence-free
state Eq.~(\ref{cd_modes}) becomes
\begin{equation*}
|\psi >_{cd}=U_{c}\left( T_{\ell }\right) U_{d}\left( T_{\ell }\right)
S_{c}\left( r\right) S_{d}\left( -r\right) |00>_{cd} \; ,
\end{equation*}
where $U_{k}\left( T_{\ell }\right) $ is the beam splitter operator
corresponding the
$k $-th mode. Consequently, the characteristic function that describes the
state $|\psi >_{cd}$ reads
\begin{equation*}
\chi _{cd}\left( \beta _{c};\beta _{d}\right) =\text{Tr}\left[ \rho
_{cd}D_{c}\left( \alpha _{c}\right) D_{d}\left( \alpha _{d}\right) \right]
\; ,
\end{equation*}
where $\rho _{cd}=|\psi >_{cd\text{ }cd}<\psi |$. The variances of
the modes $c$ and $d$ can be evaluated using the following property
of the characteristic function:
\begin{equation}
\left( -\right) ^{q}\frac{\partial ^{p+q}}{\partial \alpha _{k}^{p}\partial
\alpha _{l}^{\ast q}}\chi \left( \alpha \right) |_{\alpha,\alpha^{\ast}=0}
=\text{Tr}\left[ \rho _{cd}\left[ \left( k^{\dag }\right) ^{p}l^{q}
\right] _{\text{symmetric}}\right] \; ,
\end{equation}
with $k,l=c,d$. From this relation we can obtain the following variances:
\begin{eqnarray}
\mathrm{Var}_{X_{c}}\left( T_{\ell }\right) &=&\mathrm{Var}_{Y_{d}}\left(
T_{\ell }\right) =\frac{1-T_{\ell }\left( 1-e^{2r}\right) }{2},
\label{VarXc} \\
\mathrm{Var}_{X_{d}}\left( T_{\ell }\right) &=&\mathrm{Var}_{Y_{c}}\left(
T_{\ell }\right) =\frac{1-T_{\ell }\left( 1-e^{-2r}\right) }{2},
\label{VarXd}
\end{eqnarray}
where
\begin{equation*}
X_{k}=\frac{k+k^{\dag }}{\sqrt{2}},\text{ \ \ \ \ \ \ }Y_{k}=i\frac{-k+k^{\dag }}{\sqrt{2}}\text{\ ,}
\end{equation*}
are the quadrature operators corresponding to the $k$-th mode. Therefore,
in realistic conditions, the lower limit for the variance $\mathrm{Var}_{X_{d}}\left(
T_{\ell }\right) $ is $\left( 1-T_{\ell }\right) /2$, corresponding to
$r\rightarrow \infty $. When $T_{\ell }$ tends to $1$ (ideal case), the
variances (\ref{VarXc}) and (\ref{VarXd}) tend, respectively, to the ideal
values $e^{2r}/2$ and $e^{-2r}/2$, while the lower limit for the variance
$\mathrm{Var}_{X_{d}}\left( T_{\ell }\right) $ vanishes.

If we denote by $r^{\prime }$ the effective, actually observed
principal squeezing parameter, we have
\begin{equation*}
\mathrm{Var}_{X_{d}}\left( T_{\ell }\right) =\frac{1}{2}e^{-2r^{\prime }} \; .
\end{equation*}
The inverse relation is
\begin{eqnarray*}
r^{\prime } &=&-\frac{1}{2}\ln \left[ 2\mathrm{Var}_{X_{d}}
\left( T_{\ell}\right) \right] \\
&=&-\frac{1}{2}\ln \left[ 1-T_{\ell }\left( 1-e^{-2r}\right) \right] \; .
\end{eqnarray*}

Similar relations hold for the ancillary squeezing $s$. We may notice that
if one lets the actually observed principal squeezing parameter
$r^{\prime }$ go to zero, then the ideal squeezing parameter $r$ goes to
zero as well $\forall$ $T_{\ell }$. There is no finite value of $r>0$
and $T_{\ell }$ such that the observed squeezing vanishes. This fact
implies that decoherence can never attenuate the squeezing to vanishingly
small values.

\section{Appendix B}

\emph{Formalism of the characteristic function --}

In this appendix we describe in some detail the tunable states in terms of the
characteristic function formalism.

The state $|\zeta >_{12}|\xi >_{34}$, Eq.~(\ref{protos}), is the product of a
pair of two independent two-mode squeezing states. Therefore, the
characteristic function associated to the overall
four-mode density matrix $\rho _{1234}$ corresponding to
$|\zeta >_{12}|\xi >_{34\text{ }} $ reads as follows:
\begin{equation*}
\chi ^{\prime \prime }\left( \alpha _{12}^{\prime \prime };\alpha
_{34}^{\prime \prime }\right) =\chi _{12}\left( \alpha _{12}^{\prime \prime
}\right) \chi _{34}\left( \alpha _{34}^{\prime \prime }\right) \; ,
\end{equation*}

where

\begin{equation*}
\chi _{ij}\left( \alpha _{ij}^{^{\prime \prime }}\right) =\exp
\left\{ - \frac{1}{2}\left( \left\vert \varsigma _{i}\right\vert ^{2} +
\left\vert \varsigma _{j}\right\vert ^{2}\right) \right\} \; ,
\end{equation*}

\noindent and $\varsigma _{i,j}=\alpha _{i,j}^{\prime \prime }\cosh |\lambda
|+\alpha _{j,i}^{\prime \prime \ast }e^{i\phi _{\lambda }}\sinh |\lambda |$,
with $\lambda =\zeta $ if $i=1\wedge j=2$, and $\lambda =\xi $ if $i=3\wedge
j=4$. In order to simulate the effect of decoherence, see Fig.~(\ref{SchemeReal}),
we have introduced four thermal beam splitters $TBS$ (one for each beam), with
transmissivity $T_{th}\left( =1-R_{th}\right) $, in which each second port
is impinged by the thermal state described by the following characteristic
function:
\begin{equation*}
\chi _{k}^{th}\left( \tau _{k}^{\prime }\right) =\exp
\left\{ -\frac{1}{2} \left( 2\bar{n}_{k}^{th}+1\right)
|\tau _{k}^{\prime }|^{2}\right\} \; ,
\end{equation*}
where $\bar{n}_{th}$ is the average number of thermal quanta at
equilibrium in the $k-$th mode:
\begin{equation*}
\bar{n}_{k}^{th}=\left( e^{\hbar \omega /k_{B}T}-1\right) ^{-1} \; .
\end{equation*}

The overall characteristic function before entering the thermal beam splitters,
$\chi _{preTBS}$, describes the following eight-mode state:
\begin{equation}
\chi _{preTBS}\left( \alpha ^{\prime \prime };\tau ^{\prime }\right) =
\chi^{\prime \prime }\left( \alpha ^{\prime \prime }\right) \chi _{th}
\left( \tau ^{\prime }\right) \; ,
\label{preBSth}
\end{equation}
where $\chi _{th}\left( \tau ^{\prime }\right) = \prod\nolimits_{k=1}^{4}\chi
_{k}^{th}\left( \tau _{k}^{\prime }\right) $. The beam splitters act on the
state through a $SU(2)$ transformation that yields the following relation
among the variables of the input and output modes:
\begin{equation*}
\left\{
\begin{array}{c}
\alpha ^{\prime }=\sqrt{T_{th}}\alpha ^{\prime \prime }+\sqrt{R_{th}}\tau
^{\prime } \; , \\
\tau =\sqrt{T_{th}}\tau ^{\prime }-\sqrt{R_{th}}\alpha ^{\prime \prime } \; .
\end{array}
\right.
\end{equation*}
Therefore the input modes are related to the output modes by the following
linear transformation:
\begin{equation}
\left\{
\begin{array}{c}
\alpha ^{\prime \prime }=\sqrt{T_{th}}\alpha ^{\prime }-\sqrt{R_{th}}\tau \; ,
\\
\tau ^{\prime }=\sqrt{T_{th}}\tau +\sqrt{R_{th}}\alpha ^{\prime } \; .
\end{array}
\right.
\label{Trasfor_Tth}
\end{equation}
Using the transformations Eq.~(\ref{Trasfor_Tth}), the characteristic function
Eq.~(\ref{preBSth}) describing the state after the passage through the four thermal
beam splitters $BS_{th}$, depends on $\alpha ^{\prime }$ and $\tau $, and reads:
\begin{align}
\chi _{postTBS}\left( \alpha ^{\prime };\tau \right) & =\chi ^{\prime
\prime}\left( \sqrt{T_{th}}\alpha ^{\prime }-\sqrt{R_{th}}\tau \right)  \notag \\
& \times \chi _{th}\left( \sqrt{T_{th}}\tau +\sqrt{R_{th}}\alpha ^{\prime}\right) \; .
\end{align}
Tracing out the thermal state by putting $\tau =0$, we are left with
\begin{eqnarray}
\chi ^{\prime }\left( \alpha ^{\prime }\right) &=&\chi _{postTBS}
\left(\alpha ^{\prime };0\right)  \notag \\
&=&\chi ^{\prime \prime }\left( \sqrt{T_{th}}\alpha ^{\prime }\right)
\chi_{th}\left( \sqrt{R_{th}}\alpha ^{\prime }\right) \; .
\end{eqnarray}

The photon losses are introduced through four further beam splitters, $VBS$
($V $ for ``vacuum''), with transmissivity $T_{\ell }\left( =1-R_{\ell }\right)$,
in which each second port is occupied by a vacuum mode:
\begin{equation}
\chi _{k}^{vac}\left( \upsilon _{k}^{\prime }\right) =\exp
\left\{ -\frac{1}{2}|\upsilon _{k}^{\prime }|^{2}\right\} \; ,
\end{equation}
with $\upsilon _{k}^{\prime }$ complex coherent amplitudes. Hence, the overall
vacuum characteristic function is $\chi _{vac}\left( \upsilon ^{\prime
}\right) =\prod\nolimits_{k=1}^{4}\chi _{k}^{vac}\left( \upsilon
_{k}^{\prime }\right) $. The overall characteristic function before the
vacuum beam splitters, $\chi _{preVBS}$, reads
\begin{equation}
\chi _{preVBS}\left( \alpha ^{\prime };\upsilon ^{\prime }\right) =
\chi^{\prime }\left( \alpha ^{\prime }\right) \chi _{vac}
\left( \upsilon^{\prime } \right) \; .
\label{chi_BSvac}
\end{equation}%
In this case, the $SU(2)$ transformation defines the relations
\begin{eqnarray}
&&\left\{
\begin{array}{c}
\alpha =\sqrt{T_{\ell }}\alpha ^{\prime }+\sqrt{R_{\ell }}\upsilon^{\prime} \; , \\
\upsilon =\sqrt{T_{\ell }}\upsilon ^{\prime }-\sqrt{R_{\ell }}\alpha^{\prime } \; ,
\end{array}
\right. \\
&&\left\{
\begin{array}{c}
\alpha ^{\prime }=\sqrt{T_{\ell }}\alpha -\sqrt{R_{\ell }}\upsilon \, , \\
\upsilon ^{\prime }=\sqrt{T_{\ell }}\upsilon +\sqrt{R_{\ell }}\alpha \, .
\end{array}
\right.
\label{Traf_BSvac}
\end{eqnarray}
Thus, at the output of the beam splitters $VBS$, under the transformations
Eq.~(\ref{Traf_BSvac}), the characteristic function Eq.~(\ref{chi_BSvac}) has
evolved into
\begin{align*}
& \chi _{postVBS}\left( \alpha ;\upsilon \right) \\
& =\chi ^{\prime }\left( \sqrt{T_{\ell }}\alpha -\sqrt{R_{\ell }}\upsilon
\right) \chi _{vac}\left( \sqrt{T_{\ell }}\upsilon +\sqrt{R_{\ell }}\alpha
\right) \; .
\end{align*}
Tracing out the vacuum state ($\upsilon =0$), we are left with
\begin{align*}
\chi \left( \alpha \right) & =\chi ^{\prime }\left( \sqrt{T_{\ell }}\alpha
\right) \chi _{vac}\left( \sqrt{R_{\ell }}\alpha \right) \\
& =\chi ^{\prime \prime }\left( \sqrt{T_{th}T_{\ell }}\alpha \right) \chi
_{th}\left( \sqrt{R_{th}T_{\ell }}\alpha \right) \\
& \times \chi _{vac}\left( \sqrt{R_{\ell }}\alpha \right) \; .
\end{align*}

Considering the transformations produced by $BS_{I\text{ }}$ and $BS_{II\text{ }}$,
for the complex variables $\alpha _{1};\alpha
_{2};\alpha _{3};\alpha _{4}$, see Fig.~(\ref{SchemeReal}), we have
\begin{eqnarray*}
&&\left\{
\begin{array}{c}
\alpha _{1}=\sqrt{T_{1}}\beta _{1}-\sqrt{R_{1}}\beta _{3} \; , \\
\alpha _{3}=\sqrt{T_{1}}\beta _{3}+\sqrt{R_{1}}\beta _{1} \; ,
\end{array}%
\right. \\
&&\left\{
\begin{array}{c}
\alpha _{2}=\sqrt{T_{2}}\beta _{2}-\sqrt{R_{2}}\beta _{4} \; , \\
\alpha _{4}=\sqrt{T_{2}}\beta _{4}+\sqrt{R_{2}}\beta _{2} \; ,
\end{array}
\right.
\end{eqnarray*}
so that the four-mode characteristic function $\chi \left( \alpha \right) $
is given by
\begin{eqnarray}
&&\chi _{1234}\left( \beta _{1};\beta _{2};\beta _{3};\beta _{4}\right)
\notag \\
&=&\chi \left( \sqrt{T_{1}}\beta _{1}-\sqrt{R_{1}}\beta _{3};
\sqrt{T_{2}}\beta _{2}-\sqrt{R_{2}}\beta _{4};\right.  \notag \\
&&\left. \sqrt{T_{1}}\beta _{3}+\sqrt{R_{1}}\beta _{1};\sqrt{T_{2}}
\beta_{4}+\sqrt{R_{2}}\beta _{2}\right) \; .
\end{eqnarray}
Finally, the density matrix corresponding to the characteristic function $\chi _{1234}\left( \beta
_{1};\beta _{2};\beta _{3};\beta _{4}\right)$ is
\begin{align*}
& \rho _{1234} \\
& =\frac{1}{\pi ^{4}}\int d^{2}\beta _{1}d^{2}\beta _{2}d^{2}\beta
_{3}d^{2}\beta _{4}\chi _{1234}\left( \beta _{1};\beta _{2};\beta _{3};\beta
_{4}\right) \\
& \times D_{1}\left( -\beta _{1}\right) D_{2}\left( -\beta _{2}\right)
D_{3}\left( -\beta _{3}\right) D_{4}\left( -\beta _{4}\right) \; .
\end{align*}
At optical frequencies, the characteristic field energy $\hbar \omega $ lies
always in the range between $1.5$ and $2.5eV$, so that at room temperature $T\simeq 300K$,
the average number of thermal photons $\bar{n}_{th}$ is of the order of $10^{-30}$.
Therefore, the value of $\bar{n}_{th}$ is orders of magnitude smaller than the the mean
number of photons associated to the various quantum sources and operators. For this reason,
we have neglected throughout the thermal contribution to decoherence. In all cases, the
ideal, decoherence-free state is recovered by putting $T_{\ell }=1$.

\end{document}